\newif\ifIEEE
\IEEEfalse                    
\newif\ifLONG
\LONGtrue                     
\ifIEEE
  \documentclass[10pt,conference]{IEEEtran}
  \newcommand{\Author}[1]{#1}
  \newcommand{\Ptitle}[1]{``#1''}
\else
  \documentclass[12pt]{article}
  \usepackage{amsthm}
  \newcommand{\Author}[1]{\textsc{#1}}
  \newcommand{\Ptitle}[1]{\textsl{#1}}
\fi
\usepackage{amsfonts,bm,amssymb}
\usepackage{ifthen}
\usepackage{color}
\renewcommand{\mathbf}[1]{{\bm{#1}}}     
\newtheorem{theorem}{\indent Theorem}[section]

\newtheorem{lemma}[theorem]{\indent Lemma}
\ifIEEE
  \newcommand{\qed}{\endproof}
  \newcommand{\Endspace}{\vspace{-1ex}}  
  \newcommand{\Proof}[1]{%
                  \textit{Proof\ifthenelse{\equal{#1}{}}{}{ #1}:}}
\else
  \theoremstyle{definition}              
  \newcommand{\Proof}[1]{%
                  \textbf{Proof\ifthenelse{\equal{#1}{}}{}{ #1}.}}
  \newcommand{\Endspace}{}
\fi
\newtheorem{remark}{\indent Remark}[section]
\newenvironment{block}{%
        \begin{minipage}{\columnwidth}\vspace{1ex}
        \makebox[0ex]{}\hrulefill\makebox[0ex]{}\\*}{%
              \makebox[0ex]{}\hrulefill\makebox[0ex]{}\end{minipage}}
\newcommand{\entropy}{{\mathsf{H}}}
\newcommand{\basis}{{\mathcal{U}}}

\newcommand{\field}{{\mathbb{F}}}
\newcommand{\Naturals}{{\mathbb{N}}}
\newcommand{\GF}{{\mathrm{GF}}}
\newcommand{\code}{{\mathcal{C}}}
\newcommand{\decoder}{{\mathcal{D}}}
\newcommand{\Linear}{{\mathcal{L}}}
\newcommand{\event}{{\mathcal{E}}}
\newcommand{\Sphere}{{\mathcal{B}}}
\newcommand{\Prob}{{\mathsf{Prob}}}
\newcommand{\Expected}{{\mathsf{E}}}
\newcommand{\distance}{{\mathsf{d}}}
\newcommand{\weight}{{\mathsf{w}}}
\newcommand{\Exp}{{\mathsf{exp}}}
\newcommand{\bldc}{{\mathbf{c}}}
\newcommand{\blde}{{\mathbf{e}}}
\newcommand{\bldu}{{\mathbf{u}}}
\newcommand{\bldx}{{\mathbf{x}}}
\newcommand{\bldy}{{\mathbf{y}}}
\newcommand{\bldz}{{\mathbf{z}}}
\newcommand{\bldQ}{{\mathbf{Q}}}
\newcommand{\bldzero}{{\mathbf{0}}}
\newcommand{\Title}{On Linear Balancing Sets}
\newcommand{\Namea}{Arya Mazumdar}
\newcommand{\Addra}{Department of ECE\\
                    University of Maryland\\
                    College Park, MD 20742, USA\\
                    \texttt{arya@umd.edu}}
\newcommand{\Nameb}{Ron M. Roth}
\newcommand{\Addrb}{Computer Science Department\\
                    Technion\\
                    Haifa 32000, Israel\\
                    \texttt{ronny@cs.technion.ac.il}}
\newcommand{\Namec}{Pascal O. Vontobel}
\newcommand{\Addrc}{Hewlett--Packard Laboratories\\
                    Palo Alto, CA 94304, USA\\
                    \texttt{pascal.vontobel@ieee.org}}
\newcommand{\Footnotemark}[1]{${}^{#1}$}
\newcommand{\Footnotetext}[2]{\begin{figure}[!b]\footnotesize%
  \vspace{-3ex}\hrulefill\hfill\makebox[0em]{}\hfill\makebox[0em]{}%
  \par${}^{#1}$ #2\vspace{-0.60ex}\end{figure}\addtocounter{figure}{0}}

\begin{document}
\ifIEEE
  \title{\Title}
  \author{\IEEEauthorblockN{\Namea\Footnotemark{*}}
          \IEEEauthorblockA{\Addra}
  \and
          \IEEEauthorblockN{\Nameb\Footnotemark{*}}
          \IEEEauthorblockA{\Addrb}
  \and
          \IEEEauthorblockN{\Namec}
          \IEEEauthorblockA{\Addrc}
  }
  \maketitle
\else
  \title{\textbf{\Title}}
  \author{\textsc{\Namea\Footnotemark{*}} \\ \Addra
  \and
          \textsc{\Nameb}\Footnotemark{*} \\ \Addrb
  \and
          \textsc{\Namec} \\ \Addrc
  }
  \date{}
  \maketitle
\fi

\begin{abstract}
Let $n$ be an even positive integer and $\field$ be the field $\GF(2)$.
A word in $\field^n$ is called balanced if its Hamming weight
is $n/2$.
A subset $\code \subseteq \field^n$ is called a balancing set
if for every word $\bldy \in \field^n$ there is
a word $\bldx \in \code$ such that $\bldy + \bldx$ is balanced.
It is shown that most linear subspaces of $\field^n$ of
dimension slightly larger than $\frac{3}{2} \log_2 n$
are balancing sets. A generalization of this result
to linear subspaces that are ``almost balancing'' is also presented.
On the other hand, it is shown that the problem of deciding
whether a given set of vectors in $\field^n$ spans a balancing set,
is NP-hard. 
An application of linear balancing sets
is presented for designing efficient error-correcting coding schemes
in which the codewords are balanced.
\end{abstract}
\Footnotetext{*}{This work was done while visiting
                 the Information Theory Research Group
                 at Hewlett--Packard Laboratories,
                 Palo Alto, CA 94304, USA.}

\section{Introduction}

Let $\field$ denote the finite field $\GF(2)$
and assume hereafter that $n$ is an even positive integer.
For words (vectors) $\bldx$ and $\bldy$ in $\field^n$, denote by
$\weight(\bldx)$ the Hamming weight of $\bldx$
and by $\distance(\bldx,\bldy)$ the Hamming distance between
$\bldx$ and $\bldy$.

We say that a word $\bldz \in \field^n$ is \emph{balanced}
if $\weight(\bldz) = n/2$.
For a word $\bldx \in \field^n$, define the set
\begin{eqnarray*}
\Sphere(\bldx) & = &
\left\{ \bldx + \bldz \,:\,
\textrm{$\bldz$ is balanced} \right\} \\
& = &
\left\{ \bldy \in \field^n \,:\,
\distance(\bldy,\bldx) = n/2 \right\} \; .
\end{eqnarray*}
In particular, if $\bldzero$ denotes the all-zero word
in $\field^n$, then $\Sphere(\bldzero)$ is the set of all balanced
words in $\field^n$. It is known that
\begin{equation}
\label{eq:lowerupperbound}
\frac{2^n}{\sqrt{2n}} \le
{n \choose n/2}
= |\Sphere(\bldx)|
\le
\frac{2^n}{\sqrt{\pi n/2}}
\end{equation}
(see, for example, \cite[p.~309]{MS}).
We extend the notation $\Sphere(\cdot)$
to subsets $\code \subseteq \field^n$ by
\[
\Sphere(\code) = \bigcup_{\bldx \in \code} \Sphere(\bldx) \; .
\]

A subset $\code \subseteq \field^n$ is called
a \emph{balancing set} if $\Sphere(\code) = \field^n$;
equivalently, $\code$ is a balancing set if
for every $\bldy \in \field^n$ there exists
$\bldx \in \code$ such that
$\distance(\bldy,\bldx) = \weight(\bldy + \bldx) = n/2$
(which is also the same as saying that
for every $\bldy \in \field^n$ one has
$\Sphere(\bldy) \cap \code \ne \emptyset$).
Using the terminology of Cohen \emph{et al.} in~\cite[\S 13.1]{CHLL},
a balancing set can also be referred to
as an \emph{$\{ n/2 \}$-covering code}.

An example of a balancing set of size $n$
was presented by Knuth in~\cite{Knuth}: his set
consists of the words $\bldx_1, \bldx_2, \ldots, \bldx_n$, where
\[
\bldx_i =
\underbrace{1 1 \ldots 1}_i
\underbrace{0 0 \ldots 0}_{n{-}i} \; .
\]
It was shown by Alon \emph{et al.}
in~\cite{ABCO} that every balancing set must contain
at least $n$ words; hence, Knuth's balancing set has
the smallest possible size.

As proposed by Knuth, balancing sets can be used to efficiently
encode unconstrained binary words into balanced words as follows:
given an information word $\bldu \in \field^n$,
a word $\bldx$ in a balancing set $\code$ is found so that
$\bldu + \bldx$ is balanced.
The transmitted codeword then consists of $\bldu + \bldx$,
appended by a recursive encoding of the index
(of length $\lceil \log_2 |\code| \rceil$)
of $\bldx$ within $\code$.
Thus, when $|\code| = n$, the redundancy of the transmission is
$(\log_2 n) + O(\log \log n)$.
By~(\ref{eq:lowerupperbound}), we can get a smaller redundancy of
$\frac{1}{2} (\log_2 n) + O(1)$
using any one-to-one mapping into $\Sphere(\bldzero)$.
Such a mapping, in turn, can be implemented using enumerative coding,
but the overall time complexity will be higher than Knuth's encoder.

In many applications, the transmitted codewords are not only
required to be balanced, but also
to have some Hamming distance properties so as to provide
error-correction capabilities. Placing an error-correcting encoder
before applying any of the two balancing encoders mentioned earlier,
will generally not work, since the balancing encoder may
destroy any distance properties of its input.
One possible solution would then be to encode the raw information word
directly into a codeword of a constant-weight error-correcting code,
in which all codewords are in $\Sphere(\bldzero)$.
By a simple averaging argument one gets that
for every code $\code \subseteq \field^n$ there is at least
one word $\bldx \in \field^n$ for which the shifted set
\[
\code + \bldx =
\{ \bldy \in \field^n \,:\, \bldy - \bldx \in \code \}
\]
contains at least
$\bigl({n \choose n/2} {\bigm/} 2^n \bigr) |\code| \ge
|\code|/{\sqrt{2n}}$
balanced words.
Yet, for most known constant-weight codes,
the implementation of an encoder for such codes is
typically quite complex compared to the encoding of linear codes or
to the above-mentioned balancing methods~\cite{Sendrier}.

In this work, we will be interested in \emph{linear balancing sets},
namely, balancing sets that are linear subspaces of $\field^n$.
Our main result, to be presented in Section~\ref{sec:most},
states that most linear subspaces of $\field^n$ of dimension which is
at a (small) margin above $\frac{3}{2} \log_2 n$
are linear balancing sets. A generalization of this result to sets
which are ``almost balancing'' (in a sense to be formally defined)
will be presented in Section~\ref{sec:almostbalancing}. 
On the other hand, we will prove
(in Appendix~\ref{appendix:NP-hard}) that the problem of deciding
whether a given set of vectors in $\field^n$ spans a balancing set,
is NP-hard.

Our study of balancing sets was motivated by the potential application
of these sets in obtaining efficient coding schemes that combine
balancing and error correction, as we outline
in Section~\ref{sec:balancedcodes}.
However, we feel that linear balancing sets could be interesting
also on their own right, from a purely combinatorial point of view.

\section{Existence result}
\label{sec:exist}

    From the result in~\cite{ABCO}
we readily get the following lower bound on the dimension
of any linear balancing set.

\begin{theorem}
\label{thm:lowerbound}
\textup{\cite{ABCO}}
The dimension of every linear balancing set
$\code \subseteq \field^n$ is at least $\lceil \log_2 n \rceil$.
\end{theorem}

As mentioned earlier,
we will show that most linear subspaces of $\field^n$
of dimension slightly above $\frac{3}{2} \log_2 n$
are in fact balancing sets. We start with
the following simpler existence result,
as some components of its proof
(in particular, Lemma~\ref{lemma:one} below)
will be useful also for our random-coding result.

\begin{theorem}
\label{thm:exist}
There exists a linear balancing set in $\field^n$
of dimension $\lceil \frac{3}{2} \log_2 n\rceil$.
\end{theorem}

Theorem~\ref{thm:exist} can be seen as
the balancing-set counterpart of
the result of Goblick~\cite{Goblick}
regarding the existence of good linear covering codes
(see also Berger~\cite[pp.~201--202]{Berger}, Cohen~\cite{Cohen},
Cohen \emph{et al.}~\cite[\S 12.3]{CHLL},
and Delsarte and Piret~\cite{DP});
in fact, our proof is strongly based on their technique.
In what follows, we will adopt the formulation of~\cite{DP}.

Before proving Theorem~\ref{thm:exist}, we introduce some notation.
We denote the union $\code \cup (\code + \bldx)$
by $\code + \field \bldx$.
(When $\code$ is a linear subspace of $\field^n$
then so is $\code + \field \bldx$,
and $\code + \bldx$ is a coset of $\code$ within $\field^n$.)

We also define
\[
Q(\code) =
2^{-n} \left| \field^n \setminus \Sphere(\code) \right| =
1 - \frac{|\Sphere(\code)|}{2^n} \; .
\]
Namely, $Q(\code)$ is the probability that
$\Sphere(\bldx) \cap \code = \emptyset$,
for a randomly and uniformly selected word $\bldx \in \field^n$.

The proof of Theorem~\ref{thm:exist} makes use of the following lemma.

\begin{lemma}
\label{lemma:one}
For every subset $\code \subseteq \field^n$,
\[
2^{-n} \sum_{\bldx \in \field^n}
Q(\code + \field \bldx) = (Q(\code))^2 \; .
\]%
\Endspace
\end{lemma}

\Proof{}
The proof is essentially the first part of
the proof of Theorem~3 in~\cite{DP}, except that we replace
\ifLONG
    the Hamming sphere by $\Sphere(\cdot)$.
    For the sake of completeness, we include the proof in
    Appendix~\ref{appendix:one}.\qed
\else
    the Hamming sphere by $\Sphere(\cdot)$.\qed
\fi

\Proof{of Theorem~\ref{thm:exist}}
Again, we follow the steps of the proof of Theorem~3 in~\cite{DP}.
Write $\ell = \lceil \frac{3}{2} \log_2 n \rceil$.
We construct iteratively linear subspaces
$\code_0 \subset \code_1 \subset \cdots \subset \code_\ell$ as follows.
The subspace $\code_0$ is simply $\{ \bldzero \}$.
Given now the subspace $\code_{i-1}$, we let
\[
\code_i = \code_{i-1} + \field \bldx_i \; ,
\]
where $\bldx_i$ is a word in $\field^n$ such that
\[
Q(\code_{i-1} + \field \bldx_i) \le (Q(\code_{i-1}))^2 \; ;
\]
by Lemma~\ref{lemma:one}, such a word indeed exists.
Now,
\begin{equation}
\label{eq:upperboundonQ0}
Q(\code_0) = 1 - \frac{|\Sphere(\bldzero)|}{2^n} =
1 - 2^{-n} {n \choose n/2} \le 1 - \frac{1}{\sqrt{2n}} \; ,
\end{equation}
where the last step follows from the lower bound
in~(\ref{eq:lowerupperbound}).
Hence,
\[
Q(\code_\ell) \le (Q(\code_0))^{2^\ell} \le
\Bigl( 1 - \frac{1}{\sqrt{2n}} \Bigr)^{n^{3/2}} \le
\mathrm{e}^{-n/\sqrt{2}} < 2^{-n} \; .
\]
As $2^n Q(\code_\ell)$ is an integer, we conclude that
$Q(\code_\ell)$ is necessarily zero,
namely, $\Sphere(\code_\ell) = \field^n.$\qed

\section{Most linear subspaces are balancing sets}
\label{sec:most}

The next theorem is our main result.
Hereafter, $\Naturals$ stands for the set of natural numbers,
and the notation $\Exp(z)$ stands for
an expression of the form $a \cdot 2^{b z}$,
for some positive constants $a$ and $b$.

\begin{theorem}
\label{thm:most}
Given a function $\rho : (2 \Naturals) \rightarrow \Naturals$,
let $\code$ be a random linear subspace of $\field^n$
which is spanned by
$\lceil \frac{3}{2} \log_2 n \rceil + \rho(n)$ words
that are selected independently and uniformly from $\field^n$.
Then,
\[
\Prob
\left\{
\textrm{$\code$ is a balancing set}
\right\} \ge
1 - \Exp (-\rho(n)) \; .
\]%
\Endspace
\end{theorem}
(Thus, as long as $\rho(n)$ goes to infinity with $n$,
all but a vanishing fraction of
the ensemble of linear subspaces of $\field^n$ of dimension
$\lceil \frac{3}{2} \log_2 n \rceil + \rho(n)$
are balancing sets.)

Theorem~\ref{thm:most} is
the balancing-set counterpart of a result originally obtained by
Blinovskii~\cite{Blinovskii}, showing that most linear codes
attain the sphere-covering bound.
An alternate proof for his result
(with slightly different convergence rates as $n \rightarrow \infty$)
was then presented by Cohen \emph{et al.} in~\cite[\S 12.3]{CHLL}.
The proof that we provide for Theorem~\ref{thm:most} can be seen as
an adaptation (and refinement) of the proof of Cohen \emph{et al.}
to the balancing-set setting.

We break the proof of Theorem~\ref{thm:most} into three lemmas.
To maintain the flow of the exposition,
we will defer the proofs of the lemmas
until after the proof of Theorem~\ref{thm:most}.

\begin{lemma}
\label{lemma:two}
Let $\code_0$ be a random linear subspace of $\field^n$ which is
spanned by $\lceil \frac{1}{2} \log_2 n \rceil$ random words
that are selected independently and uniformly from $\field^n$.
There exists an absolute constant $\beta \in [0,1)$
independent of $n$ (e.g., $\beta = \frac{3}{4}$) such that
\[
\Prob \left\{ Q(\code_0) > \beta \right\} \le \Exp(-n) \; .
\]%
\Endspace
\end{lemma}

\begin{lemma}
\label{lemma:three}
Let $\code_0$ be a linear subspace of $\field^n$.
Fix a positive integer $r$, and let $\code_1$ be
a random linear subspace of $\field^n$
which is spanned by $\code_0$ and $r$ random words from $\field^n$ that
are selected uniformly and independently.
Then
\[
\Prob
\left\{
Q(\code_1) > (Q(\code_0))^{(r/2)+1}
\right\}
< (Q(\code_0))^{r/2} \; .
\]%
\Endspace
\end{lemma}

\begin{lemma}
\label{lemma:four}
Let $\code_1$ be a linear subspace of $\field^n$
and let $\code_2$ be
a random linear subspace of $\field^n$
which is spanned by $\code_1$ and
$\lceil \log_2 n \rceil$ random words from $\field^n$ that
are selected uniformly and independently.
Then
\[
\Prob
\left\{
Q(\code_2) > 0
\right\}
\le 8 Q(\code_1) \; .
\]%
\Endspace
\end{lemma}

\Proof{of Theorem~\ref{thm:most}}
It is known (e.g., from~\cite[p.~444, Theorem~9]{MS}) that
\[
\Prob \left\{
\code \ne \field^n
\right\} \le \Exp(n-\rho(n)) \; .
\]
Hence, we can assume hereafter in
the proof that $\rho(n)$ is at most linear in $n$.

Let $\basis$ be the list
of $|\basis| = \lceil \frac{3}{2} \log_2 n \rceil + \rho(n)$
random words from $\field^n$ that span $\code$,
and write $\ell = \lceil \frac{1}{2} \log_2 n \rceil$,
$t = \lceil \log_2 n \rceil$,
and $r = |\basis|{-}\ell{-}t$.
We partition the words in $\basis$
into three sub-lists,
$\basis_0$, $\basis_1$, and $\basis_2$,
of sizes $\ell$, $r$, and $t$, respectively.
We denote by $\code_0$, $\code_1$, and $\code_2$
the linear spans of
$\basis_0$,
$\basis_0 \cup \basis_1$,
and
$\basis_0 \cup \basis_1 \cup \basis_2$, respectively.

Take $\beta = \frac{3}{4}$ (say).
By Lemma~\ref{lemma:two} we get that
\begin{equation}
\label{eq:two}
\Prob \left\{ Q(\code_0) > \beta \right\} \le \Exp(-n) \; .
\end{equation}
By Lemma~\ref{lemma:three} we have
\begin{equation}
\label{eq:three}
\Prob \left\{ Q(\code_1) > \beta^{(r/2)+1} \Bigm|
Q(\code_0) \le \beta \right\} < \beta^{r/2} \; .
\end{equation}
Finally, by Lemma~\ref{lemma:four} we get
\begin{equation}
\label{eq:four}
\Prob \left\{ Q(\code_2) > 0 \Bigm|
Q(\code_1) \le \beta^{(r/2)+1} \right\}
\le (8 \beta) \cdot \beta^{r/2} \; .
\end{equation}
The result is now obtained by combining
(\ref{eq:two})--(\ref{eq:four})
and noting that $\beta^{r/2} = \Exp(-\rho(n))$.\qed

Next, we turn to the proofs of the lemmas.

\Proof{of Lemma~\ref{lemma:two}}
Write $\ell = \lceil \frac{1}{2} \log_2 n \rceil$,
and let $\bldx_1, \bldx_2, \ldots, \bldx_\ell$
denote the random words that span $\code_0$.
The proof is based on the fact that, with high probability,
the Hamming weight of each nonzero word in $\code_0$ is close to $n/2$.
Indeed, fix some nonzero vector $(a_i)_{i=1}^\ell$ in $\field^\ell$.
Then the sum $\bldx = \sum_{i=1}^\ell a_i \bldx_i$ is
uniformly distributed over $\field^n$ and, so,
by the Chernoff bound,
for every $\delta > 0$ there exists $\eta = \eta(\delta) > 0$ such that
\[
\Prob
\left\{
\bigl| \weight(\bldx) - \frac{n}{2} \bigr| > \delta n \right\}
\le 2^{-\eta n} \; .
\]
Given some $\delta \in [0,\frac{1}{2})$,
let $\event$ denote the event that $\code_0$ has dimension
(exactly) $\ell$ and each nonzero word in $\code_0$
has Hamming weight within
$\left( \frac{1}{2} \pm \delta \right) n$; namely,
\[
\event =
\Bigl\{
\textrm{%
$\displaystyle \bigl| \weight(\bldx) - \frac{n}{2} \bigr| \le \delta n$
for every
$\bldx = \sum_{i=1}^\ell a_i \bldx_i$
where $(a_i)_{i=1}^\ell \in \field^\ell \setminus \{ \bldzero \}$}
\Bigr\} \; .
\]
By the union bound we readily get that
\[
\Prob \left\{ \event \right\} >
1 - 2^\ell \cdot 2^{-\eta n} = 1 - \Exp(-n) \; .
\]

Let $\bldx$ and $\bldx'$ be two distinct words in $\code_0$,
write $\distance(\bldx,\bldx') = \tau n$,
and suppose that
$\frac{1}{2} - \delta \le \tau \le \frac{1}{2} + \delta$.
If $\tau n$ is odd then
$|\Sphere(\bldx) \cap \Sphere(\bldx')| = 0$.
Otherwise,
\begin{eqnarray}
\left| \Sphere(\bldx) \cap \Sphere(\bldx') \right|
& =  &
{\tau n \choose \tau n/2}
{(1{-}\tau) n \choose (1{-}\tau) n/2} \nonumber\\
& \le &
\frac{2^{\tau n}}{\sqrt{\pi \tau n/2}} \cdot
\frac{2^{(1-\tau) n}}{\sqrt{\pi(1{-}\tau) n/2}} \nonumber\\
& = &
\frac{2^{n+1}}{\pi n \sqrt{\tau(1{-}\tau)}} \nonumber\\
& \le &
\frac{2^{n+2}}{\pi n \sqrt{1{-}4 \delta^2}} \; ,
\label{eq:intersection1}
\end{eqnarray}
where the second step follows from the upper bound
in~(\ref{eq:lowerupperbound}).

Conditioning on the event $\event$, we get by
de Caen's lower bound~\cite{Caen} that
\begin{eqnarray}
|\Sphere(\code_0)|
& \ge &
\sum_{\bldx \in \code_0}
\frac{|\Sphere(\bldx)|^2}{
\sum_{\bldx' \in \code_0}
|\Sphere(\bldx) \cap \Sphere(\bldx')|} \nonumber \\
& > &
2^\ell {n \choose n/2}^2
\biggm/
\left(
\frac{2^\ell \cdot 2^{n+2}}{\pi n \sqrt{1{-}4 \delta^2}} +
{n \choose n/2} \right) \nonumber \\
\label{eq:lowerboundsphere}
& \ge &
2^n
\biggm/
\biggl( \frac{8}{\pi \sqrt{1{-}4 \delta^2}} +
\frac{\sqrt{2 n}}{2^\ell} \biggr) \; ,
\end{eqnarray}
where in the last step we have used
the lower bound in~(\ref{eq:lowerupperbound}).
On the other hand, we also have $2^\ell \ge \sqrt{n}$ and, so, writing
\[
\beta(\delta) =
1 - \Bigl( \frac{8}{\pi \sqrt{1{-}4 \delta^2}} + \sqrt{2} \Bigr)^{-1}
\; ,
\]
we get that, conditioned on the event $\event$,
\begin{equation}
\label{eq:beta}
Q(\code_0) = 1 - \frac{|\Sphere(\code_0)|}{2^n} \le \beta(\delta) \; .
\end{equation}
The result follows by recalling that
$\Prob \left\{ \event \right\} \ge 1 - \Exp(-n)$
and observing that $\beta(\delta) < 1$
for every $\delta \in [0,\frac{1}{2})$
(in particular, there is some $\delta$ for which
$\beta(\delta) = \frac{3}{4} > \beta(0)$).\qed

\begin{remark}
\label{remark:simplex1}
Suppose that $\code_0(m,\ell)$ is
an $\ell$-dimensional linear subspace of
the linear $[n{=}2^m,m,2^{m-1}]$ code over $\field$ obtained
by appending a fixed zero coordinate to every codeword
of the binary $[2^m{-}1,m,2^{m-1}]$ simplex code.
In this case, we can substitute $\delta = 0$
in~(\ref{eq:beta}) and obtain that
$Q(\code_0(m,\ell)) \le \beta(0) \approx 0.748$,
for every $\ell$ in the range $m/2 \le \ell \le m$.
Thus, $\code_0(m,\ell)$
can replace the random code $\code_0$ in Lemma~\ref{lemma:two}.
If $\ell$ grows sufficiently fast with $m$ so that
$\ell{-}(m/2)$ tends to infinity,
then from~(\ref{eq:lowerboundsphere}) it follows that
\[
\lim_{m,\, \ell{-}(m/2) \rightarrow \infty}
Q(\code_0(m,\ell)) \le 1 - \frac{\pi}{8} \approx 0.607 \; .
\]
Let $\code'_0 = \code'_0(m,\ell)$ be given by
$\code_0(m,\ell) + \field \bldx$, where
$\bldx$ is an odd-weight word in $\field^n$.
For $m > 1$ we have
$|\Sphere(\code'_0)| = 2 |\Sphere(\code_0(m,\ell))|$.
Therefore, when $m,\, \ell{-}(m/2) \rightarrow \infty$,
we can bound
$Q(\code'_0)$ from above by $1 - (\pi/4) \approx 0.215$.\qed
\end{remark}

\Proof{of Lemma~\ref{lemma:three}}
Let $\bldx_1, \bldx_2, \ldots, \bldx_r$ be
the random words that, together with
$\code_0$, span (the random code) $\code_1$.
Obviously,
$\Sphere(\code_0 + \bldx_i) \subseteq \Sphere(\code_1)$
and $Q(\code_0 + \bldx_i) = Q(\code_0)$
for every $i = 1, 2, \ldots, r$.
Hence, the expected value of $Q(\code_1)$
(taken over all the independently and uniformly distributed
words $\bldx_1, \bldx_2, \ldots, \bldx_r \in \field^n$) satisfies
\begin{eqnarray*}
\Expected \left\{ Q(\code_1) \right\}
& = &
2^{-n} \sum_{\bldy \in \field^n}
\Prob \left\{ \bldy \not\in \Sphere(\code_1) \right\} \\
& \le &
2^{-n} \sum_{\bldy \in \field^n \setminus \Sphere(\code_0)}
\prod_{i=1}^r
\Prob \left\{ \bldy \not\in \Sphere(\code_0 + \bldx_i) \right\} \\
& = &
(Q(\code_0))^{r+1} \; .
\end{eqnarray*}
Therefore,
\begin{eqnarray*}
\lefteqn{
\Prob \left\{ Q(\code_1) > (Q(\code_0))^{(r/2)+1} \right\}
} \makebox[2ex]{} \\
& \le &
\Prob \left\{ Q(\code_1) > (Q(\code_0))^{-r/2}
\Expected \left\{ Q(\code_1) \right\} \right\} \\
& < &
(Q(\code_0))^{r/2} \; ,
\end{eqnarray*}
where the last step follows from Markov's inequality.\qed

\Proof{of Lemma~\ref{lemma:four}}
The result is obvious when
$Q(\code_1) \not\in (0,\frac{1}{8})$; so we assume
hereafter in the proof that $Q(\code_1)$ is within
that interval.
Write $t = \lceil \log_2 n \rceil$,
and let $\bldx_1, \bldx_2, \ldots, \bldx_t$ be
the random words that, together with
$\code_1$, span $\code_2$.
For $i = 0, 1, 2, \ldots, t$, define the linear space
$\Linear_i$ iteratively by $\Linear_0 = \code_1$ and
\[
\Linear_i = \Linear_{i-1} + \field \bldx_i \; .
\]

Letting $\bldQ_i$ stand for (the random variable) $Q(\Linear_i)$
and $\omega_i$ for $2^i/(8 Q(\code_1))$,
by Lemma~\ref{lemma:one} and Markov's inequality
we get for every $i = 1, 2, \ldots, t$ that,
conditioned on an instance of $\Linear_{i-1}$,
\begin{eqnarray*}
\lefteqn{
\Prob \left\{
\bldQ_i > \bldQ_{i-1}^2 \omega_i
\Bigm|
\Linear_{i-1}
\right\}
} \makebox[2ex]{} \\
& = &
\Prob \Bigl\{
Q(\Linear_{i-1} + \field \bldx_i) > \bldQ_{i-1}^2 \omega_i
\Bigm|
\Linear_{i-1}
\Bigr\} \\
& \le &
\frac{1}{\omega_i} = (8 Q(\code_1)) \cdot 2^{-i} \; .
\end{eqnarray*}
Hence, for every $i = 1, 2, \ldots, t$,
\begin{eqnarray*}
\lefteqn{
\Prob \Bigl\{
\bldQ_t > \bldQ_0^{2^t}
\prod_{i=1}^t \omega_i^{2^{t-i}}
\Bigr\}
} \makebox[2ex]{} \\
& \le &
\Prob \Bigl\{
\bigcup_{i=1}^t
\bigl( \bldQ_i > \bldQ_{i-1}^2 \omega_i \bigr)
\Bigr\} \\
& \le &
\sum_{i=1}^t
\Prob \left\{
\bldQ_i > \bldQ_{i-1}^2 \omega_i
\right\} \\
& \le &
\sum_{i=1}^t \frac{1}{\omega_i} < 8 Q(\code_1) \; .
\end{eqnarray*}
Substituting
$\bldQ_0 = Q(\code_1)$ and $\bldQ_t = Q(\code_2)$, we conclude that
\[
\Prob \Bigl\{
Q(\code_2) > (Q(\code_1))^{2^t}
\prod_{i=1}^t \omega_i^{2^{t-i}}
\Bigr\}
< 8 Q(\code_1) \; ,
\]
where
\begin{eqnarray*}
(Q(\code_1))^{2^t}
\prod_{i=1}^t \omega_i^{2^{t-i}}
\!\!\!
& < &
(Q(\code_1))^{2^t}
\Bigl( \prod_{i=1}^\infty \omega_i^{2^{-i}} \Bigr)^{2^t} \\
& = &
(Q(\code_1))^{2^t}
\left(
\frac{2^{\sum_{i=1}^\infty i 2^{-i}}}
{(8 Q(\code_1))^{\sum_{i=1}^\infty 2^{-i}}}
\right)^{2^t} \\
& = &
2^{-2^t} \le 2^{-n} \; .
\end{eqnarray*}
The result follows by recalling that
the events ``$Q(\code_2) \ge 2^{-n}$'' and
``$Q(\code_2) > 0$'' are identical.\qed

Figure~\ref{fig:balancingsets} lists the generator matrices of
linear $[n,k,d]$ codes over $\field$ that form linear balancing sets,
for several values of $n$ that are divisible by $4$.
These matrices were found using a greedy algorithm and
they do not necessarily generate the smallest sets,
except for $n = 12$ and $n = 20$, where the sets
attain the lower bound of Theorem~\ref{thm:lowerbound}
(in addition, for the case $n = 20$,
the set attains the Griesmer bound~\cite[\S 17.5]{MS}).

\begin{figure}[hbt]
\begin{block}
\centering
\newenvironment{Matrix}[1]{\scriptsize\arraycolsep0.20ex%
                           \renewcommand{\arraystretch}{0.8}%
                           $\left(\begin{array}{#1}}
                           {\end{array}\right)$}
\renewcommand{\arraystretch}{0.40}
\begin{tabular}{lc}
& \\
$[8,3,3]:$ &
\begin{Matrix}{cccccccc}
0&0&0&0&1&1&1&1\\
0&1&1&1&0&0&1&0\\
1&0&0&0&1&1&0&0
\end{Matrix}
\\ & \\
$[12,4,5]:$ &
\begin{Matrix}{cccccccccccc}
0&0&0&0&0&0&1&1&1&1&1&1\\
0&0&0&1&1&1&0&0&1&1&1&0\\
1&0&1&0&0&1&0&1&1&1&0&0\\
1&1&1&1&0&0&0&0&1&0&0&0
\end{Matrix}
\\ & \\
$[16,5,7]:$ &
\begin{Matrix}{cccccccccccccccc}
0&0&0&0&0&0&0&0&1&1&1&1&1&1&1&1\\
0&0&0&1&1&1&1&1&0&0&0&0&1&1&1&0\\
0&1&1&0&0&1&1&1&0&1&1&1&1&1&0&0\\
1&1&0&1&0&1&1&0&1&1&0&0&1&0&0&0\\
1&1&1&1&1&1&1&1&0&0&0&1&0&0&0&0
\end{Matrix}
\\ & \\
$[20,5,9]:$ &
\begin{Matrix}{cccccccccccccccccccc}
0&0&0&0&0&0&0&0&0&0&1&1&1&1&1&1&1&1&1&1\\
0&0&0&0&0&1&1&1&1&1&0&0&0&0&1&1&1&1&1&0\\
0&1&1&1&1&0&0&1&1&1&0&1&1&1&0&0&1&1&0&0\\
1&1&1&0&0&1&0&1&1&0&1&1&0&0&1&0&1&0&0&0\\
1&1&0&1&0&1&1&1&0&1&0&0&1&0&0&1&0&0&0&0
\end{Matrix}
\\ & \\
$[24,6,9]:$ &
\begin{Matrix}{cccccccccccccccccccccccc}
0&0&0&0&0&0&0&0&0&0&0&0&1&1&1&1&1&1&1&1&1&1&1&1\\
0&0&0&0&0&0&0&1&1&1&1&1&0&0&0&0&0&0&1&1&1&1&1&0\\
0&0&0&1&1&1&1&0&0&1&1&1&0&0&1&1&1&1&0&1&1&1&0&0\\
0&0&1&0&0&1&1&1&1&0&1&1&1&1&0&0&1&1&1&1&1&0&0&0\\
1&1&1&1&1&1&1&1&0&1&0&0&0&0&0&0&0&0&0&1&0&0&0&0\\
1&1&0&1&0&1&0&1&1&0&1&0&1&0&0&0&0&0&1&0&0&0&0&0
\end{Matrix}
\\ & \\
$[28,6,11]:$ &
\begin{Matrix}{cccccccccccccccccccccccccccc}
0&0&0&0&0&0&0&0&0&0&0&0&0&0&1&1&1&1&1&1&1&1&1&1&1&1&1&1\\
0&0&0&0&0&0&0&1&1&1&1&1&1&1&0&0&0&0&0&0&1&1&1&1&1&1&1&0\\
0&0&0&1&1&1&1&0&0&0&1&1&1&1&0&0&1&1&1&1&0&0&1&1&1&1&0&0\\
0&0&1&0&0&1&1&0&1&1&0&0&1&1&1&1&0&0&1&1&1&1&1&1&1&0&0&0\\
1&1&1&1&1&1&0&1&1&0&1&1&1&0&0&0&0&0&0&0&0&0&0&1&0&0&0&0\\
1&0&1&1&0&1&1&1&0&1&1&0&0&1&1&0&0&0&0&0&0&0&1&0&0&0&0&0
\end{Matrix}
\\ & \\
$[32,7,13]:$ &
\begin{Matrix}{cccccccccccccccccccccccccccccccc}
0&0&0&0&0&0&0&0&0&0&0&0&0&0&0&0&1&1&1&1&1&1&1&1&1&1&1&1&1&1&1&1\\
0&0&0&0&0&0&0&0&0&1&1&1&1&1&1&1&0&0&0&0&0&0&0&0&1&1&1&1&1&1&1&0\\
0&0&0&0&0&1&1&1&1&0&0&0&0&0&1&1&0&0&0&0&0&1&1&1&0&1&1&1&1&1&0&0\\
0&1&1&1&0&0&1&1&0&0&1&0&0&1&0&1&0&1&0&1&1&1&0&1&1&0&1&0&1&0&0&0\\
0&1&1&0&1&1&1&0&1&0&0&1&0&1&0&1&1&1&1&1&0&1&1&0&1&0&1&1&0&0&0&0\\
1&0&1&1&0&1&1&1&0&0&0&1&1&1&0&0&0&1&1&0&1&1&1&0&0&1&1&0&0&0&0&0\\
1&0&1&0&0&1&0&0&1&0&0&0&0&0&0&1&1&1&0&1&1&1&1&1&0&1&0&0&0&0&0&0
\end{Matrix}
\\ & \\
\end{tabular}

\end{block}
\caption{Bases of linear balancing sets for $n = 8, 12,16, \ldots, 32$.}
\label{fig:balancingsets}
\end{figure}

\begin{remark}
\label{remark:simplex3}
In view of Remark~\ref{remark:simplex1},
when $n = 2^m$ (or, more generally, when $n$ is ``close'' to $2^m$),
Theorem~\ref{thm:most} holds also for
the smaller ensemble where we fix
$\lceil m/2 \rceil$ basis elements of
the random code $\code$ to be linearly independent codewords
of the code $\code_0(m,\lceil m/2 \rceil)$
defined in Remark~\ref{remark:simplex1}.
Furthermore, if these $\lceil m/2 \rceil$ rows
are replaced by $\ell$ basis elements of
the code $\code_0'(m,\ell)$ (as defined in that remark),
then the value $\beta$ in the proof
of Theorem~\ref{thm:most} can be taken as
$1 - (\pi/4) \; (\approx 0.215)$
whenever $\ell{-}(m/2)$ goes to infinity
(yet more slowly than $\rho(n)$).\qed
\end{remark}

We leave it open to find an explicit
construction of linear balancing sets in $\field^n$
of dimension $O(\log n)$.
We also mention the following intractability result.

\begin{theorem}
\label{thm:NP-hard}
Given as input a basis of a linear subspace $\code$ of $\field^n$,
the problem of deciding whether $\code$ is a balancing set, is NP-hard.
\end{theorem}

The proof of Theorem~\ref{thm:NP-hard} is obtained
by some modification of the reduction in~\cite{McLoughlin} from
\textsc{Three-Dimensional Matching}.
\ifLONG
    We include the proof in Appendix~\ref{appendix:NP-hard}.
\else
    We omit the details due to space limitations.
\fi

\section{Linear almost-balancing sets}
\label{sec:almostbalancing}

While the code $\code_0(m,\ell{=}m)$
in Remark~\ref{remark:simplex1} is such that
$Q(\code_0(m,m))$ is bounded away from zero,
this code can be seen as ``almost balancing'' in the following sense:
for every word $\bldy \in \field^n$ (where $n = 2^m$)
there exists a codeword $\bldx \in \code_0(m,m)$ such that
$\left| \distance(\bldy,\bldx) - (n/2) \right| \le \sqrt{n}/2$.
The proof of this fact is similar to the one showing that
the covering radius of the first-order Reed--Muller code
is at most $(n - \sqrt{n})/2$~\cite[pp.~241--242]{CHLL}
(specifically, in the line following Eq.~(9.2.4) therein,
simply reverse the inequality in
``$| \langle \cdot,\cdot \rangle| \ge \sqrt{n}$'';
see also~(\ref{eq:sumofsq}) below).

Next, we formalize the notion of almost balancing sets
and present generalizations for Theorems~\ref{thm:exist}
and~\ref{thm:most}.
In what follows, we fix some function
$\lambda : 2\Naturals \rightarrow \Naturals$
such that $\lambda(n) < n/2$,
and write $\lambda = \lambda(n)$ for simplicity.
For a word $\bldx \in \field^n$ define the set
\[
\Sphere_{\lambda}(\bldx) =
\left\{ \bldy \in \field^n \,:\,
|\distance(\bldy,\bldx) - n/2| \le \lambda \right\} \; .
\]
As was the case for $\lambda = 0$,
the notation $\Sphere_{\lambda}(\cdot)$
can be extended to subsets $\code \subseteq \field^n$ by
\[
\Sphere_{\lambda}(\code) =
\bigcup_{\bldx \in \code} \Sphere_{\lambda}(\bldx) \; .
\]

A subset $\code \subseteq \field^n$ is called
a \emph{$\lambda$--almost-balancing set}
if $\Sphere_{\lambda}(\code) = \field^n$;
equivalently, $\code$ is a $\lambda$--almost-balancing set if
for every $\bldy \in \field^n$ there exists
$\bldx \in \code$ such that
$|\distance(\bldy,\bldx) - n/2| \le \lambda$.

The following theorem can be seen as
a generalization of Theorem \ref{thm:exist}.

\begin{theorem}
\label{thm:exists_almost}
Suppose that $\lambda = \lambda(n) = O(\sqrt{n})$.
There exists a linear $\lambda$--almost-balancing set of dimension
$\left\lceil \frac{3}{2}\log_2{n} - \log_2{(2\lambda+1)}+
O(\lambda^2/n) \right\rceil$.
\end{theorem}

\Proof{}
We follow the steps of the proof of Theorem \ref{thm:exist},
with $Q(\code_i)$ replaced by a term
$Q_\lambda(\code_i)$ which equals $1 - 2^{-n} \Sphere_\lambda(\code_i)$,
and with~(\ref{eq:upperboundonQ0}) replaced by
an upper bound on
$Q_\lambda(\code_0) = Q_\lambda(\{ \bldzero \})$
which we shall now derive.

Let $\entropy : [0,1] \rightarrow [0,1]$ be the binary entropy function
$\entropy(z) = -(z \log_2 z) - (1{-}z) \log_2(1{-}z)$. Then,
\begin{eqnarray}
|\Sphere_{\lambda}(\bldzero)|
& = &
\sum_{i=(n/2)-\lambda}^{(n/2)+\lambda}{n \choose i} \nonumber \\
& \ge &
(2\lambda+1) {n \choose n/2 - \lambda} \nonumber \\
& \ge &
\frac{2\lambda+1}{\sqrt{2n (1 -4 (\lambda/n)^2)}} \cdot
2^{n \entropy (\frac12 - \frac{\lambda}{n} )} \nonumber \\
& \ge &
\label{eq:lowerlambdasphere}
\frac{2\lambda+1}{\sqrt{2n }} \cdot
2^{n \entropy (\frac12 - \frac{\lambda}{n} )} \; ,
\end{eqnarray}
where the penultimate step follows from
a well known lower bound on binomial
coefficients~\cite[p.~309]{MS}. From~(\ref{eq:lowerlambdasphere})
we have,
\[
Q_\lambda(\code_0) \le
1 - \frac{2\lambda+1}{\sqrt{2n}} \cdot
2^{-n ( 1- \entropy (\frac12 - \frac{\lambda}{n} ) )} \; ,
\]
thereby obtaining the counterpart of~(\ref{eq:upperboundonQ0}).
Proceeding as in the proof Theorem \ref{thm:exist},
we see that
$\left\lceil \frac{3}{2}\log_2{n} -
\log_2{(2\lambda+1)}+n\left(1-\entropy\left(\frac12 -
\frac{\lambda}{n}\right)\right)\right\rceil$
basis elements are sufficient to span
a linear $\lambda$--almost-balancing set.

Finally, using the Taylor series expansion for $\entropy(\frac12- z)$
and recalling that $\lambda = O(\sqrt{n})$, we obtain
\begin{eqnarray}
n \Bigl( 1 - \entropy \Bigl(\frac12 -\frac{\lambda}{n}\Bigr) \Bigr)
& = &
2\frac{\lambda^2}{n} + \frac43\frac{\lambda^4}{n^3} + \ldots \nonumber\\
& = &
\label{eq:taylorentr}
\frac{\lambda^2}{n} \Bigl( 2 + o(1) \Bigr)
= O \Bigl( \frac{\lambda^2}{n} \Bigr) \; ,
\end{eqnarray}
thereby completing the proof.\qed

Observe that for $n = 2^m$ and $\lambda = \lfloor \sqrt{n}/2 \rfloor$,
the code $\code_0(m,m)$ realizes the dimension guaranteed
in Theorem~\ref{thm:exists_almost}.

The following theorem is a generalization of Theorem \ref{thm:most}.

\begin{theorem}
\label{thm:most_almost}
Suppose that $\lambda = \lambda(n) = O(\sqrt{n})$.
Given a function $\rho : 2\Naturals \rightarrow \Naturals$,
let $\code$ be a random linear subspace of $\field^n$ that is spanned by
$\left\lceil \frac{3}{2}\log_2{n} -
\log_2{(2\lambda+1)}\right\rceil+ \rho(n)$ words
selected independently and uniformly from $\field^n$.
Then,
\[
\Prob
\left\{
\textrm{$\code$ is a $\lambda$--almost-balancing set}
\right\} \ge
1 - \Exp (-\rho(n)) \; .
\]%
\Endspace
\end{theorem}

\Proof{}
The proof is the same as that of Theorem \ref{thm:most},
except that $Q(\cdot)$ is replaced by $Q_\lambda(\cdot)$
in Lemmas~\ref{lemma:three} and~\ref{lemma:four} (and in their proofs),
and Lemma~\ref{lemma:two} is replaced by the following lemma.\qed

\begin{lemma}
\label{lemma:two_almost}
Suppose that $\lambda = O(\sqrt{n})$,
and let $\code_0$ be a random linear subspace of $\field^n$ which is
spanned by
$\lceil \frac{1}{2} \log_2 n - \log_2{(2\lambda+1)}\rceil$
random words
that are selected independently and uniformly from $\field^n$.
There exists an absolute constant $\beta \in [0,1)$ such that
\[
\Prob \left\{ Q_\lambda(\code_0) > \beta \right\} \le \Exp(-n) \; .
\]%
\Endspace
\end{lemma}

The proof of Lemma~\ref{lemma:two_almost}
\ifLONG
    can be found in Appendix~\ref{appendix:two_almost}.
\else
    is omitted due to space limitations.
\fi

While Theorems~\ref{thm:exists_almost}
and~\ref{thm:most_almost} only cover
the case where $\lambda = O(\sqrt{n})$, we next show that when
$\lambda = \Omega(\sqrt{n})$, it is fairly easy to
obtain an explicit construction for
linear $\lambda$--almost-balancing sets
with relatively small dimensions.
Specifically, let $s$ and $m$ be any two positive integers,
and set $n = s \cdot 2^m$ and
$\lambda = \lfloor \sqrt{sn}/2 \rfloor$.
The construction described below yields
a linear $\lambda$--almost-balancing set of dimension
at most $2 (\log_2{n} - \log_2{(2\lambda)})$.

Given $m$ and $s$, let $\code_0 = \code_0(m,m)$ be
the linear $[M{=}2^m,m,2^{m-1}]$ code over $\field$
as in Remark~\ref{remark:simplex1},
and let $\bldc_1, \bldc_2, \ldots, \bldc_M$ denote the codewords
of $\code_0$.
It is shown in~\cite{CHLL} that for every word $\bldy \in \field^M$,
\begin{equation}
\label{eq:sumofsq}
\sum_{i=1}^M
\left( M - 2 \distance(\bldy,\bldc_i) \right)^2 = M^2
\end{equation}
(from which one gets that there exists at least one codeword
$\bldc_i \in \code_0$ such that
$|(M/2) - \distance(\bldy,\bldc_i) | \le \sqrt{M}/2$;
see the discussion at the beginning of this section).

Consider now the code $\code_0^{(s)}$ which consists of the words
$\bldx_1, \bldx_2, \ldots, \bldx_M$, where
\[
\bldx_i =
\underbrace{(\bldc_i \,|\, \bldc_i \,|\, \ldots \,|\, \bldc_i)}_%
            {s \; \mathrm{times}} \; ,
\quad
i = 1, 2, \ldots, M \; .
\]
Clearly, $\code_0^{(s)}$ is a linear $[n{=}sM,m]$ code over $\field$.
Given a word $\bldy \in \field^n$,
we write it as $(\bldy_1 \,|\, \bldy_2 \,|\, \ldots \,|\, \bldy_s)$
where each block $\bldy_j$ is in $\field^M$, and define
\[
z_{i,j} = M - 2 \distance(\bldy_j,\bldc_i) \; ,
\quad
i = 1, 2, \ldots, M \; , \quad
j = 1, 2, \ldots, s \; .
\]
Obviously,
\[
n - 2 \distance(\bldy,\bldx_i) = \sum_{j=1}^s z_{i,j}
\; , \quad
i = 1, 2, \ldots, M \; ,
\]
and, so,
\begin{eqnarray*}
\sum_{i=1}^M
\Bigl(  n - 2 \distance(\bldy,\bldx_i) \Bigr)^2
& = &
\sum_{i=1}^M \Bigl( \sum_{j=1}^s z_{i,j} \Bigr)^2 \\
& \le &
s \sum_{i=1}^M
\sum_{j=1}^s z_{i,j}^2 \\
& = &
s \sum_{j=1}^s  \sum_{i=1}^M z_{i,j}^2
\; \stackrel{(\ref{eq:sumofsq})}{=} \;
s^2 M^2 \; ,
\end{eqnarray*}
where the inequality follows from the convexity of $z \mapsto z^2$.
Hence, there is at least one index
$i \in \{ 1, 2, \ldots, M \}$ for which
\[
|n - 2 \distance(\bldy,\bldx_i)| \le s \sqrt{M} = \sqrt{s n} \; .
\]
We conclude that $\code_0^{(s)}$ is
a linear $\lambda$--almost-balancing set with
$\lambda = \lfloor \sqrt{s n}/2 \rfloor$, and its dimension is
$m = \log_2 {(n/s)} \le 2 (\log_2{n} - \log_2{(2\lambda)})$.

We end this section by comparing our results to
the following generalization of Theorem~\ref{thm:lowerbound}.

\begin{theorem}
\label{thm:lowerbound_almost}
\textup{\cite{ABCO}}
The dimension of every linear $\lambda$--almost-balancing set
$\code \subseteq \field^n$ is at least
$\lceil\log_2{n} -\log_2(2\lambda+1)\rceil$.
\end{theorem}

For $\lambda = O(\sqrt{n})$, there is still an additive gap of
approximately $\frac12\log_2{n}$ between the lower bound
and the upper bound guaranteed by
Theorem \ref{thm:exists_almost},
and for $\lambda = \Omega(\sqrt{n})$,
the dimension of $\code_0^{(s)}$ is approximately twice the lower bound.

\section{Balanced error-correcting codes}
\label{sec:balancedcodes}

In this section, we consider a potential application of
linear balancing sets in designing an efficient coding scheme
that maps information words into balanced words
that belong to a linear error-correcting code;
as such, the scheme combines error-correction capabilities with
the balancing property.

The underlying idea is as follows. Let $\code$ be
a linear $[n,k,d]$ code over $\field$ with the length $n$
and minimum distance $d$ chosen so as to satisfy
the required correction capabilities. Suppose, in addition, that
we can write $\code$ as a direct sum of two linear subspaces
$\code'$ and $\code''$ of dimensions $k'$ and $k''$, respectively,
\begin{equation}
\label{eq:decomposition}
\code = \code' \oplus \code'' =
\left\{
\bldc + \bldx \,:\,
\bldc \in \code', \bldx \in \code''
\right\}
\; ,
\end{equation}
where $\code''$ is a balancing set\footnote{%
For the scheme to work, it actually suffices that words in $\code''$
balance only the elements of $\code'$,
rather than all the words in $\field^n$.}.
Now, if $k''$ is ``small'' (which means that $k'$ is close to $k$),
we can encode by first mapping
a $k'$-bit information word $\bldu$ into a codeword
$\bldc \in \code'$, and then finding a word $\bldx \in \code''$
so that $\bldc + \bldx$ is balanced.
The transmitted codeword is then the (balanced) sum
$\bldc + \bldx$. The mapping $\bldu \mapsto \bldc$ can be
implemented simply as a linear transformation,
whereas the balancing word $\bldx$ can be found
by exhaustively searching over the $2^{k''}$ elements of $\code''$.
At the receiving end, we apply a decoder for $\code$
(for correcting up to $(d{-}1)/2)$ errors) to
a (possibly noisy) received word $\bldc + \bldx + \blde$,
where $\blde$ is the error word.
Clearly, if $\weight(\blde) \le (d{-}1)/2$, we will be able
to recover $\bldc + \bldx$ successfully, thereby retrieving~$\bldu$.

Obviously, such as scheme is useful only when $k''$ is indeed small:
first, $k''$ affects the effective rate
(given by $k'/n =  (k{-}k'')/n$) and, secondly,
the encoding process---as described---is exponential in $k''$.
Yet, not always is there a decomposition of $\code$
as in~(\ref{eq:decomposition})
that results in a small dimension $k''$ of $\code''$
(in fact, for some codes $\code$, such a composition does not exist
at all).

A possible solution would then be to reverse the design process
and start by first selecting the code $\code'$
so that it has the desired rate $R = k'/n$ and
a ``slightly'' higher minimum distance $d'$ than
the desired value $d$. In addition, we assume
that there is an efficient (i.e., polynomial-time)
decoding algorithm $\decoder'$ for $\code'$
that corrects any pattern of up to $(d{-}1)/2$ errors.

Next, we select $\code''$ to be a random linear code spanned by
$k'' = \lceil \frac{3}{2} \log_2 n \rceil + \rho(n)$ words
that are chosen independently and uniformly from $\field^n$,
for some function $\rho(n) = o(\log n)$ that grows to infinity.
By Theorem~\ref{thm:most}, the code $\code''$ will be a balancing
set with probability $1 - \Exp(-\rho(n)) = 1 - o(1)$,
and the choice of $k''$ guarantees
that an exhaustive search for the balancing word $\bldx$
during encoding will take $O(n^{3/2 + \epsilon})$ iterations,
for an arbitrarily small $\epsilon > 0$
(if the search fails---an event that may occur
with probability $o(1)$---we can simply replace the code $\code''$).
The receiving end can
be informed of the choice of the code $\code''$ by, say, using
pseudo-randomness instead of randomness (and flagging a skip
when failing to find a balancing word $\bldx$).

It remains to consider the distance properties of the direct sum
$\code = \code' \oplus \code''$; specifically, we need
the subset of balanced words in $\code$ to have minimum distance
at least $d$; in particular, every balanced word in $\code$
should have a unique decomposition of the form
$\bldc + \bldx$ where $\bldc \in \code'$ and $\bldx \in \code''$.
When this condition holds, the decoding can proceed
as follows. Given a received word $\bldy \in \field^n$,
we enumerate over all words $\bldx \in \code''$ and
then apply the decoder $\decoder'$ to each difference $\bldy-\bldx$.
Decoding will be successful if the number of errors did not
exceed $(d{-}1)/2$, and the decoding complexity will be
$O(n^{3/2+\epsilon})$ times the complexity of $\decoder'$.

The next lemma considers the case where the code $\code'$
lies below the Gilbert--Varshamov bound.
Hereafter, $V(n,t)$ stands for $\sum_{i=0}^t {n \choose i}$.

\begin{lemma}
\label{lemma:gv}
Suppose that $\code'$ is
a linear $[n,k',d']$ code over $\field$ that satisfies
$2^{k'} \cdot V(n,d'{-}1) \le 2^n$.
For every $d \le d'$,
the minimum distance $\distance(\cdot)$ of
(the random code) $\code = \code' \oplus \code''$ satisfies
\[
\Prob
\left\{
\distance(\code) < d
\right\}
< 2^{k''} \cdot \frac{V(n,d{-}1)}{V(n,d'{-}1)} \; .
\]%
\Endspace
\end{lemma}

\Proof{}
The code $\code$ contains $|\code| - |\code'|$ random codewords,
each being uniformly distributed over $\field^n$
and therefore each having probability
$V(n,d{-}1)/2^n$ to be of Hamming weight less than $d$.
The result follows from the union bound.\qed

It is well known (see~\cite[p.~310]{MS})
that for any integer $t = \theta n \le n/2$,
\[
\frac{1}{\sqrt{2n}} \cdot 2^{n \entropy(\theta)}
\le V(n,t) \le
2^{n \entropy(\theta)}
\; ,
\]
where $\entropy : [0,1] \rightarrow [0,1]$ is the binary entropy
function defined earlier.
Hence, taking $k'' \le (\frac{3}{2} + \epsilon) \log_2 n$,
we get from Lemma~\ref{lemma:gv} and the concavity
of $z \mapsto \entropy(z)$ that
\[
\Prob
\left\{
\distance(\code) < d
\right\}
< \sqrt{2} \cdot n^{2+\epsilon}
\Bigl( \frac{d'{-}1}{n{-}d'{+}1} \Bigr)^{d'-d}
\; .
\]

Thus, to achieve a vanishing probability,
$\Prob \left\{ \distance(\code) < d \right\}$,
of ending up with a ``bad'' code $\code$ as $n$ goes to infinity,
it suffices to take $d' = d + O(\log n)$ when
$d/n$ is fixed and bounded away from zero,
or $d'= d+ O(1)$ when $d$ is fixed.

\begin{remark}
\label{eq:iterativedecoding}
Instead of a decoding process whereby we enumerate
over the codewords of $\code''$ and then apply the decoder
$\decoder'$, we could use a decoder for the whole
direct sum $\code$, if techniques such as iterative decoding are
applicable to $\code$: in such circumstances, the advantage of
the linearity of $\code$ is apparent.
Linearity certainly helps if we are interested only
in error detection rather than full correction,
in which case the decoding amounts to just computing
a syndrome with respect to any parity-check matrix of $\code$.\qed
\end{remark}

\ifLONG
\ifIEEE\appendices\else\appendix\section*{\centering{Appendices}}\fi

\section{Proof of Lemma~\ref{lemma:one}}
\label{appendix:one}

\Proof{}
We have,
\begin{eqnarray*}
|\Sphere(\code + \field \bldx)|
& = &
|\Sphere(\code \cup (\code + \bldx))| \\
& = &
|\Sphere(\code)| + |\Sphere(\code + \bldx)| -
|\Sphere(\code) \cap \Sphere(\code + \bldx)| \\
& = &
2 |\Sphere(\code)| -
|\Sphere(\code) \cap (\Sphere(\code) + \bldx)| \; .
\end{eqnarray*}
Hence,
\[
\sum_{\bldx \in \field^n} |\Sphere(\code + \field \bldx)| =
2^{n+1} |\Sphere(\code)| - \sum_{\bldx \in \field^n}
|\Sphere(\code) \cap (\Sphere(\code) + \bldx)| \; .
\]
Now,
\begin{eqnarray*}
\lefteqn{
\sum_{\bldx \in \field^n}
|\Sphere(\code) \cap (\Sphere(\code) +\bldx)|
} \makebox[5ex]{} \\
& = &
|\{(\bldx,\bldy) : \bldx \in \field^n, \bldy \in \Sphere(\code),
\bldy \in \Sphere(\code) + \bldx\}|\\
& = &
|\{(\bldx,\bldy) : \bldx \in \field^n, \bldy \in \Sphere(\code),
\bldx \in \Sphere(\code) + \bldy\}|\\
& = &
|\Sphere(\code)|^2 \; .
\end{eqnarray*}
Therefore,
\[
\sum_{\bldx \in \field^n} |\Sphere(\code + \field \bldx)| =
2^{n+1} |\Sphere(\code)| - |\Sphere(\code)|^2 \; .
\]
Using the definition of $Q(\cdot)$ the lemma is proved.\qed

\section{Proof of Theorem~\ref{thm:NP-hard}}
\label{appendix:NP-hard}

\newcommand{\graph}{{\mathcal{G}}}
\newcommand{\matching}{{\mathcal{M}}}
\newcommand{\blds}{{\mathbf{s}}}
\newcommand{\bldh}{{\mathbf{h}}}
\newcommand{\bldone}{{\mathbf{1}}}

We prove Theorem~\ref{thm:NP-hard} below,
starting by recalling the reduction that is used
in~\cite{McLoughlin} to show the intractability of
computing the covering radius of a linear code.

Let $\graph = (V_1{:}V_2{:}V_3,E)$ be a tripartite hyper-graph
with a vertex set which is the union of the disjoint sets
$V_1$, $V_2$, and $V_3$ of the same size $t$,
and a hyper-edge set
$E = \{ e_1, e_2, \ldots, e_m \} \subseteq V_1 \times V_2 \times V_3$.

The reduction in~\cite{McLoughlin} maps $\graph$
into a $3t \times 8m$ parity-check matrix
$H = H_\graph = ( \, H_e \, )_{e \in E}$,
where each block $H_e$ is a $3t \times 8$ matrix over $\field$
whose rows and columns are indexed by
$u \in V_1 \cup V_2 \cup V_3$
and $(a_1 \, a_2 \, a_3) \in \field^3$, respectively,
and is computed from the hyper-edge
$e = (v_{e,1},v_{e,2},v_{e,3})$ as follows:
\[
(H_e)_{u,(a_1 a_2 a_3)} =
\left\{
\begin{array}{lcl}
0       && \textrm{if $u \ne v_{e,\ell}$ for $\ell = 1, 2, 3$} \\
a_\ell  && \textrm{if $u = v_{e,\ell}$}
\end{array}
\right.
\; .
\]
(Namely, the three nonzero rows in $H_e$ are indexed by
the vertices that are incident with the hyper-edge $e$, and
these rows form a $3 \times 8$ matrix whose columns range over all
the elements of $\field^3$.)

A \emph{matching} in $\graph$ is a subset
$\matching \subseteq E$ of size $t$ such that
no two hyper-edges in $\matching$ are incident with the same
vertex (thus, every vertex of $\graph$ is incident with
exactly one hyper-edge in $\matching$).

For our purposes, we can assume that every vertex in $\graph$
is incident with at least one hyper-edge
(or else no matching exists).
Under these conditions, $m \ge t$ and the matrix $H$ has full rank
(since it contains the identity matrix of order $3t$).

The proof in~\cite{McLoughlin} is based on the following two facts:
\begin{list}{}{\settowidth{\labelwidth}{(ii)}}
\item[(i)]
There is a matching $\matching$ in $\graph$ if and only if
the all-one column vector $\bldone$ in $\field^{3t}$ can be written
as a sum of (exactly) $t$ columns of $H$
(note that $\bldone$ cannot be the sum of less than $t$ columns).
Those columns then must be those that are indexed by
$(1\,1\,1)$ in all blocks $H_e$ such that $e \in \matching$.
\item[(ii)]
If $\matching$ is a matching in $\graph$ then
every column vector in $\field^{3t}$ can be written as
a sum $\sum_{e \in \matching} \bldh_e$,
where each $\bldh_e$ is a column in $H_e$.
\end{list}

Let $\code = \code_\graph$ be the linear $[8m,8m{-}3t]$ code
over $\field$ with a parity-check matrix $H$.
It readily follows from facts~(i) and~(ii) that $\graph$
has a matching if and only if every coset of
$\code$ within $\field^{8m}$ has a word of Hamming weight $t$.

    From facts~(i)--(ii) we get the following lemma.

\begin{lemma}
\label{lemma:NP-hard}
Suppose that $t > 1$ and that $\graph$ contains a matching.
Then every column vector in $\field^{3t}$
can be obtained as a sum of $w$ distinct columns in $H$, for
every $w$ in the range $t \le w \le 8m{-}t$.
\end{lemma}

\Proof{}
Let $\matching$ be a matching which is assumed to exist in $\graph$.
Given $w \in \{ t, t{+}1, \ldots, 8m{-}t \}$, write
\[
\sigma = \min \{ 8(m{-}t), w{-}t \} \; ,
\]
and let $\bldx$ be a column vector in $\field^{3t}$
which is the sum of $\sigma$ columns in $H$ that do \emph{not} belong
to the $t$ blocks $H_e$ that correspond to $e \in \matching$.
Also, write
\[
\tau = w - \sigma =
\left\{
\begin{array}{lcl}
t            && \textrm{if $w \le 8m{-}7t$} \\
w - 8(m{-}t) && \textrm{otherwise}
\end{array}
\right.
\; ,
\]
and note that $t \le \tau \le 7t$.

Given an arbitrary column vector $\blds \in \field^{3t}$,
we show that there are $w$ distinct columns in $H$ that sum to $\blds$.
By fact~(ii), for every $e \in \matching$
there is a column $\bldh_e$ in $H_e$ such that
\begin{equation}
\label{eq:syndrome}
\blds = \bldx + \sum_{e \in \matching} \bldh_e \; .
\end{equation}
Furthermore, it follows from the structure of each block $H_e$
that when $\bldh_e \ne \bldzero$, then for every
integer $r$ in the range $1 \le r \le 7$ there exist
$r$ distinct columns
$\bldh_{e,1}, \bldh_{e,2}, \ldots, \bldh_{e,r}$ in $H_e$ such that
\[
\bldh_e = \sum_{j=1}^r \bldh_{e,j} \; .
\]
The same holds also when $\bldh_e = \bldzero$ for values of $r$
in $\{ 0, 1, 3, 4, 5, 7, 8 \}$.

We conclude that we can find $t$ nonnegative integers
$(r_e)_{e \in \matching}$
such that the following two conditions hold:
\begin{itemize}
\item
$\sum_{e \in \matching} r_e = \tau \; (\in \{ t, t{+}1, \ldots, 7t \})$,
and---
\item
For each $e \in \matching$, the column vector $\bldh_e$ can be
written as a sum of (exactly) $r_e$ distinct columns of $H_e$.
\end{itemize}
Thus, the right-hand side of~(\ref{eq:syndrome})
can be expressed as a sum of
$\sigma + \sum_{e \in \matching} r_e = \sigma + \tau = w$
distinct columns in $H$.\qed

\Proof{of Theorem~\ref{thm:NP-hard}}
Given a hyper-graph $\graph$,
consider the linear $[16m{-}2t,8m{-}3t]$ code
$\code'_\graph$ over $\field$
with an $(8m{+}t) \times (16m{-}2t)$ parity-check matrix
\[
H' = H'_\graph =
\left(
\begin{array}{c|c}
0 & I \\
\hline
H & 0
\end{array}
\right)
\; ,
\]
where $H = H_\graph$ and $I$ is the identity matrix of order $8m{-}2t$.
Next, we show that there is a matching in $\graph$ if and only if
every coset of $\code'_\graph$ contains a balanced word
(i.e., a word of Hamming weight $8m{-}t$).

Suppose that $\graph$ contains a matching $\matching$.
We show that every column vector
$\blds \in \field^{8m+t}$ can be expressed as a sum of
(exactly) $8m{-}t$ distinct columns of $H'$.
Write $\blds^T = ( \blds_1^T \,|\, \blds_2^T )$,
where $\blds_1$ consists of the first $8m{-}2t$ entries of $\blds$
(and $\blds_2$ consists of the remaining $3t$ entries).
By Lemma~\ref{lemma:NP-hard},
there exist $w = 8m{-}t{-}\weight(\blds_1)$
distinct columns in $H$ that sum to $\blds_2$.
Hence, by the structure of $H'$ it follows that $H'$
contains $w + \weight(\blds_1) = 8m{-}t$ columns that sum to $\blds$.

Conversely, suppose that
every coset of $\code'_\graph$ contains a balanced word.
In particular, this means that the all-one vector
in $\field^{8m+t}$ can be expressed as a sum of $8m{-}t$ columns
of $H'$. Now, the last $8m{-}2t$ columns of $H'$ must
be included in this sum;
this, in turn, implies that the all-one vector
$\bldone$ in $\field^{3t}$ can be written as a sum of
$t$ columns of $H$. The result follows from fact~(i).\qed

\section{Proof of Lemma~\ref{lemma:two_almost}}
\label{appendix:two_almost}

\Proof{}
We will follow along the steps of the proof of Lemma~\ref{lemma:two},
except that~(\ref{eq:intersection1}) needs to be replaced by
a different upper bound which we now derive.
Given some $\delta \in [0,\frac{1}{2})$,
let $\bldx$ and $\bldx'$ be two distinct words in $\code_0$
with $\distance(\bldx,\bldx') = \tau n$ where
$\frac{1}{2} - \delta \le \tau \le \frac{1}{2} + \delta$.
The number of words $\bldy \in \field^n$ such that
$\distance(\bldx,\bldy) = i$ and
$\distance(\bldx',\bldy) = j$ is given by
\[
p_{i,j}^{(\tau n)} =
{\tau n \choose (j{-}i{+}\tau n)/2}%
                              {(1{-}\tau) n \choose (i{+}j{-}\tau n)/2}
\]
(here we assume that the binomial coefficient
${m \choose k}$ is equal to $0$ unless $m$ and $k$ are both
nonnegative integers and $m\ge k$). Hence,
\[
\left| \Sphere_{\lambda}(\bldx) \cap \Sphere_{\lambda}(\bldx') \right|
\le
\sum_{i=(n/2)-\lambda}^{(n/2)+\lambda}
\sum_{j=(n/2)-\lambda}^{(n/2)+\lambda} p_{i,j}^{(\tau n)} \; .
\]
It can be easily verified that when $\tau n$ is even then
\[
\max_{i,j}p_{i,j}^{(\tau n)} = {\tau n \choose \tau n/2}
{(1{-}\tau) n \choose (1{-}\tau) n/2}
\stackrel{(\ref{eq:lowerupperbound})}{\le} 
\frac{2^{\tau n}}{\sqrt{\pi \tau n/2}} \cdot
\frac{2^{(1-\tau) n}}{\sqrt{\pi(1{-}\tau) n/2}} \; ,
\]
and when $\tau n$ is odd then
\begin{eqnarray*}
\max_{i,j}p_{i,j}^{(\tau n)}
& = &
{\tau n \choose (\tau n + 1)/2}
{(1{-}\tau) n \choose ((1{-}\tau) n + 1)/2} \\
& = &
\frac{1}{4}
{\tau n + 1\choose (\tau n + 1)/2}
{(1{-}\tau) n + 1 \choose ((1{-}\tau) n + 1)/2} \\
& \stackrel{(\ref{eq:lowerupperbound})}{\le} &
\frac{2^{\tau n}}{\sqrt{\pi \tau n/2}} \cdot
\frac{2^{(1-\tau) n}}{\sqrt{\pi(1{-}\tau) n/2}} \; .
\end{eqnarray*}
In either case we have:
\begin{eqnarray}
\left| \Sphere_{\lambda}(\bldx) \cap \Sphere_{\lambda}(\bldx') \right|
& \le &
(2\lambda + 1)^2
\max_{i,j}p_{i,j}^{(\tau n)} \nonumber \\
& \le &
(2\lambda + 1)^2
\frac{2^{n+1}}{\pi n \sqrt{\tau(1{-}\tau)}} \nonumber \\
\label{eq:intersection1_almost}
& \le &
(2\lambda + 1)^2\frac{2^{n+2}}{\pi n \sqrt{1{-}4 \delta^2}} \; .
\end{eqnarray}
In addition, from (\ref{eq:lowerlambdasphere})
and~(\ref{eq:taylorentr}) we get:
\begin{equation}
\label{eq:lowerlambdasphere1almost}
|\Sphere_{\lambda}(\bldx)| \ge
\frac{2\lambda+1}{\sqrt{2n }} \cdot 2^{n-O(1)} \; .
\end{equation}
We now proceed as in the proof of Lemma~\ref{lemma:two},
with~(\ref{eq:intersection1_almost}) replacing~(\ref{eq:intersection1})
and with~(\ref{eq:lowerlambdasphere1almost}) replacing the lower bound
in~(\ref{eq:lowerupperbound}): by de Caen's lower bound~\cite{Caen}
we get a bound which is similar to~(\ref{eq:lowerboundsphere}),
in which we plug
$\ell = \lceil \frac{1}{2} \log_2 n - \log_2{(2\lambda+1)}\rceil$.
The result follows.\qed
\else\fi


\begin{thebibliography}{99}
\bibitem{ABCO}
    \Author{N. Alon, E.E. Bergmann, D. Coppersmith, A.M. Odlyzko,}
    \Ptitle{Balancing sets of vectors,}
    \textit{IEEE Trans.\ Inform.\ Theory,} 34 (1988), 128--130.
\bibitem{Berger}
    \Author{T. Berger,}
    \textit{Rate Distortion Theory,}
    Prentice-Hall, Englewood Cliffs, New Jersey, 1971.
\bibitem{Blinovskii}
    \Author{V.M. Blinovskii,}
    \Ptitle{Covering the Hamming space with sets translated by
    linear code vectors,}
    \textit{Probl.\ Inform.\ Transm.,} 26 (1990), 196--201.
\bibitem{Caen}
    \Author{D. de Caen,}
    \Ptitle{A lower bound on the probability of a union,}
    \textit{Disc.\ Math.,} 169 (1997), 217--220.
\bibitem{Cohen}
    \Author{G. Cohen,}
    \Ptitle{A nonconstructive upper bound on covering radius,}
    \textit{IEEE Trans.\ Inform.\ Theory,} 29 (1983), 352--353.
\bibitem{CHLL}
    \Author{G. Cohen, I. Honkala, S. Litsyn, A. Lobstein,}
    \textit{Covering Codes,}
    North-Holland, Amsterdam, 1997.
\bibitem{DP}
    \Author{P. Delsarte, P. Piret,}
    \Ptitle{Do most binary linear codes achieve the Goblick bound
    on the covering radius?,}
    \textit{IEEE Trans.\ Inform.\ Theory,} 32 (1986), 826--828.
\bibitem{Goblick}
    \Author{T.J. Goblick, Jr.,}
    \textit{Coding for a discrete information source with a distortion
    measure,}
    Ph.D. dissertation,
    Department of Electrical Engineering,
    Massachusetts Institute of Technology,
    Cambridge, Massachusetts, 1962.
\bibitem{Knuth}
    \Author{D.E. Knuth,}
    \Ptitle{Efficient balanced codes,}
    \textit{IEEE Trans.\ Inform.\ Theory,} 32 (1986), 51--53.
\bibitem{MS}
    \Author{F.J. MacWilliams, N.J.A. Sloane,}
    \textit{The Theory of Error-Correcting Codes,}
    North-Holland, Amsterdam, 1977.
\bibitem{McLoughlin}
    \Author{A.M. McLoughlin,}
    \Ptitle{The complexity of computing the covering radius of a code,}
    \textit{IEEE Trans.\ Inform.\ Theory,} 30 (1984), 800--804.
\bibitem{Sendrier}
    \Author{N. Sendrier,}
    \Ptitle{Encoding information into constant weight words,}
    \textit{Proc.\ 2005 Int'l Symp.\ Inform.\ Theory (ISIT~2005),}
    Adelaide, Australia (2005), 435--438.
\end{thebibliography}
\end{document}